\documentclass[]{tADP2e}

\usepackage{graphicx,graphics,color,epsfig} 
\usepackage{bm}
\usepackage{amsmath}
\usepackage{amssymb}

\usepackage{type1cm}
\usepackage{graphicx}
\usepackage{array}

\def\bc{\begin{center}}
\def\ec{\end{center}}

\def\beq{\begin{equation}}
\def\eeq{\end{equation}}

\begin{document}
\title{Properties of Graphene: A Theoretical Perspective}

\author{D.S.L. Abergel$^{\rm a}$, V. Apalkov$^{\rm b}$, J. Berashevich$^{\rm a}$, 
K. Ziegler$^{\rm c}$, and Tapash Chakraborty$^{\rm a}$$^{\ast}$
\thanks{$^\ast$Corresponding author. Email: tapash@physics.umanitoba.ca 
\vspace{6pt}}\\\vspace{6pt} $^{\rm a}${\em{Department of Physics and 
Astronomy, University of Manitoba, Winnipeg, Canada R3T 2N2}}; 
$^{\rm b}${\em{Department of Physics and Astronomy, Georgia State University, 
Atlanta, Georgia 30303, USA}}; $^{\rm c}${\em{Institut f\"ur Physik,
Universit\"at Augsburg, D-86135 Augsburg, Germany}}
}

\maketitle

\begin{abstract}
The electronic properties of graphene, a two-dimensional crystal of carbon atoms, 
are exceptionally novel. For instance the low-energy quasiparticles in graphene behave 
as massless chiral Dirac fermions which has led to the experimental observation of many
interesting effects similar to those predicted in the relativistic regime.
Graphene also has immense potential to be a key ingredient of new 
devices such as single molecule gas sensors, ballistic transistors, and spintronic 
devices. Bilayer graphene, which consists of two stacked monolayers and where the
quasiparticles are massive chiral fermions, has a quadratic low-energy band
structure which generates very different scattering properties from those of the
monolayer. It also presents the unique property that a tunable band gap can be opened
and controlled easily by a top gate. These properties have made bilayer graphene
a subject of intense interest.

In this review, we provide an in-depth description of the physics of monolayer
and bilayer graphene from a theorist's perspective. We discuss the physical
properties of graphene in an external magnetic field, reflecting the chiral
nature of the quasiparticles near the Dirac point with a Landau level at zero
energy. We address the unique integer quantum Hall effects, the role of electron 
correlations, and the recent observation of the fractional quantum Hall effect in 
the monolayer graphene. The quantum Hall effect in bilayer graphene is fundamentally
different from that of a monolayer, reflecting the unique band structure of
this system. The theory of transport in the absence of an external magnetic field
is discussed in detail, along with the role of disorder studied in various
theoretical models. We highlight the differences and similarities between monolayer 
and bilayer graphene, and focus on thermodynamic properties such as the compressibility,
the plasmon spectra, the weak localization correction, quantum Hall effect,
and optical properties.

Confinement of electrons in graphene is nontrivial due to Klein tunneling.
We review various theoretical and experimental studies of quantum confined structures 
made from graphene. The band structure of graphene nanoribbons and the role of the 
sublattice symmetry, edge geometry and the size of the nanoribbon on the electronic 
and magnetic properties are very active areas of research, and a detailed review of 
these topics is presented. Also, the effects of substrate interactions, adsorbed atoms,
lattice defects and doping on the band structure of finite-sized graphene
systems are discussed. We also include a brief description of graphane -- 
gapped material obtained from graphene by attaching hydrogen atoms to 
each carbon atom in the lattice.
\end{abstract}

\vspace*{1.0cm}
\begin{center}
\bf{Contents}
\end{center}
\begin{tabbing}

1. \= XX \= XXXXXXXXXXXXXXXXXXXXXXXXXXXXXXXXXXXXX
\= XXXXXX \= XXXX \kill
\> \> \> \> {\sc page}\\
1. \> Introduction \> \> \> 3\\
\> 1.1. A sheet of molecular chicken wire \> \> \> 6\\
\> 1.2. Massless Dirac fermions \> \> \> 9\\
\> 1.3. How it's made \> \> \> 10\\
\> 1.4. Graphene devices \> \> \> 11\\
2. Graphene in a magnetic field \> \> \> \> 12\\
\> 2.1. Landau levels in graphene \> \> \> 12\\
\> 2.2. Anomalous quantum Hall effect \> \> \> 17\\
\> \> 2.2.1. Experimental observation of the quantum Hall effect in graphene \> \> 17\\
\> \> 2.2.2. Symmetry breaking: theoretical models \> \> 21\\
\> \> 2.2.3. Symmetry breaking: disorder effects \> \> 22\\
\> \> 2.2.4. Symmetry breaking: electron-electron interaction effects \> \> 24\\
\> \> 2.2.5. Symmetry breaking: lattice distortion \> \> 28\\
\> \> 2.2.6. Edge states in a strong magnetic field\> \> 29\\
\> 2.3. Fractional quantum Hall effect \> \> \> 32\\
3. Bilayer graphene \> \> \> \> 35\\
\> 3.1. Sample fabrication and identification \> \> \> 37\\
\> \> 3.1.1. Optical identification of exfoliated bilayer graphene \> \> 37\\
\> \> 3.1.2. Atomic force microscopy, and miscellaneous diagnostic techniques \> \> 38\\
\> \> 3.1.3. Raman spectroscopy \> \> 40\\
\> 3.2. Tight-binding model \> \> \> 41\\
\> \> 3.2.1. Nearest neighbor and next-nearest neighbor models \> \> 41\\
\> \> 3.2.2. Trigonal warping \> \> 46\\
\> \> 3.2.3. Effective low-energy theory \> \> 47\\
\> 3.3. Band gap in bilayer graphene \> \> \> 48\\
\> \> 3.3.1. Band gap in the tight-binding model \> \> 48\\
\> \> 3.3.2. Experimental evidence of gap \> \> 51\\
\> \> 3.3.3. Ab-initio simulations \> \> 54\\
\> 3.4. Quantum Hall effect \> \> \> 56\\
\> \> 3.4.1. Experimental picture \> \> 56\\
\> \> 3.4.2. Tight-binding description of low-energy Landau levels \> \> 57\\
\> \> 3.4.3. Magneto-optical properties of bilayer graphene \> \> 60\\
\> \> 3.4.4. The effect of trigonal warping on the Landau level spectrum \> \> 63\\
\> \> 3.4.5. Electron-electron interactions in the zero-mode Landau levels \> \> 63\\
\> 3.5. Electron-electron interactions in bilayer graphene \> \> \> 65\\
\> 3.6. Phonon anomalies and electron-phonon coupling \> \> \> 68\\
\> 3.7. Device proposals utilizing bilayer graphene \> \> \> 70\\
4. Many-body and optical properties of graphene \> \> \> \> 71\\
\> 4.1. Electronic compressibility \> \> \> 72\\
\> \> 4.1.1. Monolayer graphene \> \> 72\\
\> \> 4.1.2. Bilayer graphene \> \> 75\\
\> 4.2. Plasmon dispersion in graphene \> \> \> 77\\
\> \> 4.2.1. Monolayer graphene \> \> 77\\
\> \> 4.2.2. Bilayer graphene \> \> 80\\
\> 4.3. Graphene in a strong electromagnetic field \> \> \> 83\\
5. Zero-field transport in graphene \> \> \> \> 85\\
\> 5.1. Basic experimental facts \> \> \> 86\\
\> 5.2. Tight-binding model and low-energy approximation \> \> \> 87\\
\> \> 5.2.1. Random fluctuations \> \> 90\\
\> \> 5.2.2. Density of states \> \> 91\\
\> 5.3. Theory of transport \> \> \> 92\\
\> \> 5.3.1. Boltzmann approach \> \> 92\\
\> \> 5.3.2. Kubo formalism \> \> 94\\
\> \> 5.3.3. Semiclassical approach: phenomenological scattering rate \> \> 96\\
\> 5.4. Perturbation theory for disorder \> \> \> 97\\
\> 5.5. Numerical simulations \> \> \> 99\\
\> 5.6. Field theory of diffusion: the nonlinear sigma model \> \> \> 100\\
\> \> 5.6.1. Saddle-point approximation \> \> 100\\
\> \> 5.6.2. Two-particle Green's function and diffusion\> \> 101\\
\> \> 5.6.3. Scaling relation of the two-particle Green's function\> \> 102\\
\> \> 5.6.4. Diffusion coefficients\> \> 103\\
\> \> 5.6.5. Scattering rate $\eta$ for the gapless case \> \> 103\\
\> 5.7. Metal-insulator transition \> \> \> 105\\
6. Confinement of electrons in graphene \> \> \> \> 107\\
\> 6.1. Quantum dots in graphene islands \> \> \> 108\\
\> 6.2. Electron trapping in graphene quantum dots \> \> \> 116\\
\> 6.3. Quantum dots with sharp boundaries \> \> \> 120\\
\> 6.4. Quantum dots in a magnetic field: Numerical studies \> \> \> 122\\
\> 6.5. Magnetic quantum dots \> \> \> 125\\
\> 6.6. Confinement of massive relativistic electrons in graphene \> \> \> 127\\
\> 6.7. Quantum dots in bilayer graphene \> \> \> 128\\
7. Localized states at the edges of graphene nanoribbons \> \> \> \> 129\\
\> 7.1. Localization of the electron density at the edges \> \> \> 130\\
\> 7.2. Experimental evidence for localized edge states \> \> \> 132\\
\> 7.3. Stabilization of the edge states \> \> \> 133\\
\> \> 7.3.1. The nearest-neighbor interactions \> \> 133\\
\> \> 7.3.2. Coulomb interactions \> \> 135\\
\> 7.4. Spin ordering, symmetry and band gap \> \> \> 136\\
\> 7.5. Band gap: confinement effect and edge shape \> \> \> 138\\
\> 7.6. Graphene nanoribbons in an electric field \> \> \> 143\\
\> 7.7. Nanoscale graphene \> \> \> 146\\
\> 7.8. Bilayer graphene nanoribbons and the effects of edges \> \> \> 147\\
8. Manipulation of band gap and magnetic properties of graphene \> \> \> \> 152\\
\> 8.1. Interaction of graphene with a substrate \> \> \> 153\\
\> 8.2. Doping of graphene through adsorption \> \> \> 158\\
\> \> 8.2.1. Adsorption of non-metals on graphene: Experimental results \> \> 158\\
\> \> 8.2.2. Adsorption of non-metals on graphene: Theoretical approaches \> \> 159\\
\> \> 8.2.3. From graphene to graphane \> \> 165\\
\> \> 8.2.4. Adsorption of metal atoms on graphene: Experimental results \> \> 168\\
\> \> 8.2.5. Adsorption of metal atoms on graphene: Theoretical approaches \> \> 170\\
\> 8.3. Lattice defects \> \> \> 174\\
\> \> 8.3.1. Vacancy defects  \> \> 175\\
\> \> 8.3.2. Vacancy defects saturated by hydrogen \> \> 177\\
\> \> 8.3.3. Divacancy defects \> \> 178\\
\> \> 8.3.4. Crystallographic and chemisorption defects \> \> 179\\
\> \> 8.3.5. Substitutional doping of graphene \> \> 180\\
\> 8.4. Functionalization of the edges \> \> \> 181\\
Acknowledgements \> \> \> \> 186\\
References \> \> \> \> 186\\[2ex]
\end{tabbing}

\voffset=1.5truecm

\section{Introduction}

Everything about graphite involves a mix of very old and very young.  Known to
man since ancient times (ca.~1500CE), graphite is as ubiquitous as the {\it
lead} in a pencil, and yet the subject of our current review, graphene, being
a single atomic layer of graphite, was isolated only in 2004! That discovery
marked the beginning of the academic equivalent of a gold rush which has
become a major topic of research for the condensed matter and materials
physics community, along with chemists, electrical engineers, and device
specialists.  Several thousand papers have been written in the past couple of
years that have attempted to explain every aspect of the electronic properties
of graphene. There are review articles, long and short (see, for example,
\cite{review_Nov,review_allen,review_Ando,review_Gus,review_Per}), special journal 
issues \cite{special_issues} and popular magazine articles (see, for example,
\cite{popular}). This development at `Mozartian speed' is primarily due to the
fact that a two-dimensional system of electrons in graphene behaves rather
uniquely as compared to its counterpart in semiconductor systems.  In fact,
many of the fundamental properties of graphene that were crucial for the
present developments, were already reported in the early part of the past
century, merely waiting to be confirmed experimentally until now.

In graphene, one finds a new class of low-dimensional system, only one atom
thick, with vast potential for applications in future nanotechnology.  Our
review is organized as follows.  In this section, we introduce graphene by
describing its crystal structure, and discussing its band structure via the
frequently-used tight-binding model. We also discuss the low-energy properties
of this material, and in particular we focus on the linear (Dirac-like) nature
of the energy dispersion near the edges of the Brillouin zone, and on the
chiral nature of the low-energy electrons. We also briefly discuss fabrication
techniques for graphene, and whet the appetite for study of this material by
describing some of the devices utilizing the unique properties of graphene
which have already been created in the laboratory.

Section 2 deals with the quantum Hall effect, i.e., quantization of Hall
conductance as a function of the magnetic field or the electron density, that
was initially discovered in conventional nonrelativistic two-dimensional
electron systems. The effect is a direct manifestation of Landau quantization
of electron dynamics. An electron system in graphene, being a two-dimensional
system, also shows Landau quantization of electron motion and the
corresponding quantum Hall effect, which has been observed experimentally. The
relativistic massless nature of the energy dispersion law in graphene results
in striking differences between the quantum Hall effect observed in
graphene and in conventional two-dimensional systems. In graphene,  quantum
Hall effect can be
observed even at room temperature, while in nonrelativistic systems it is
observable only at low temperatures. The quantized Hall effect in graphene
occurs not at integer values as in the conventional Hall effect, but at
half-integer values.  Such anomalous behavior of the quantum Hall effect is
due to massless relativistic nature of the charge carrier dispersion and the
electron-hole symmetry of the system. In addition to anomalous half-integer
values of the Hall conductance a rich structure of Hall plateaus has been
observed experimentally. This structure is associated with the lifting of valley
and spin degeneracy of the Landau levels.  Different many-body
mechanisms of lifting of the degeneracy of the Landau levels have been
proposed in the literature. These mechanisms are reviewed in detail. The
specific features of the many-particle excitations of the quantum Hall states,
the fractional quantum Hall effect in graphene and the unique structure of the
quantum Hall state edge states are also discussed.

In Section 3, we discuss specific aspects of bilayer graphene, and try to
highlight the similarities and differences between this and the monolayer
material. We introduce experimental techniques for distinguishing the number
of layers in a graphene flake.  We present the tight-binding formalism in
order to derive the quadratic low energy spectrum, and to discuss the
influence of trigonal warping and the formation of a band gap. We describe the
quantum Hall effect and the formation of the zero energy level with doubled
degeneracy, which is unique to this system. The interactions between electrons
are fascinating in this material, and several properties are distinct from
both the monolayer and traditional two-dimensional electron systems,
and we describe the formation of spin-polarized and other ordered states. The
interactions between electrons and phonons are also important (for example, in
the context of Raman scattering experiments) so we briefly describe the phonon
anomalies and the the electron-phonon interaction. 
Lastly, we review some of the proposals for devices which utilize bilayer
graphene in their design. 

Electronic properties that are intimately related to electron-electron
interactions, viz., the compressibility and plasmon dispersion in a
two-dimensional electron gas show unique behavior in graphene.
The compressibility of a two-dimensional electron gas is an important physical
quantity that is deduced from the ground state energy. It provides important
information about the electron correlations, the chemical potential, and the
stability of the system, etc. In Section~4, we discuss
the unique behavior of the
electron compressibility in monolayer and bilayer graphene.  In this section, we also
describe the excitation spectra of graphene in the presence of the spin-orbit
interaction within the random-phase approximation. The spin-orbit
interaction opens a
gap between the valence and conduction bands and between the intraband and
inter-band electron-hole excitation continuum of the semimetal Dirac system.
As a result, one sees a dramatic change in the long-wavelength dielectric
function of the system. An undamped plasmon mode appears in the electron-hole
continuum (EHC) gap reflecting the interplay between the intraband and
interband electron correlations.  In undoped bilayer graphene, the static
screening effect is anisotropic and much stronger than that in monolayer
graphene. The dynamic screening shows the properties of a Dirac gas in the low
frequency limit and of Fermi gas in the high frequency limit. A similar
transition from the Dirac gas to the Fermi gas is also observed in the plasmon
spectrum. In doped bilayer graphene, the plasmon spectrum is quite similar to
that of Fermi gas for momentum less than half the Fermi momentum while
becoming softer at higher momentum. We close this section with a discussion of
the properties of graphene in a strong external electromagnetic field. The
possibility of inducing valley-polarized currents by irradiating gapped
bilayer graphene is described.

In Section 5, we review the transport behavior of monolayer and bilayer
graphene in the absence of an external magnetic field, focusing on properties
in the vicinity of the charge neutrality points. Beginning with the classical
Boltzmann approach we compare the latter with the more general linear-response
(Kubo) approach. The effect of electron-electron and electron-phonon
interaction as well as the effect of different types of disorder is discussed.
Of all these effects disorder seems to be the most important one. We present
and compare several schemes of approximation for disorder averaging, including
a semiclassical approximation, perturbation theory, saddle-point approximation
and nonlinear-sigma model calculations.  Finally, the properties of a random
gap and a related metal-insulator transition is investigated.

Quantum dots or {\it artificial atoms} \cite{Chakraborty_99,Chakraborty_92}
are crucial building blocks in many nanoscale semiconductor applications.
Their unique properties, such as superior transport and tunable optical
spectra, are originated from their zero dimensionality, which results in
discrete energy spectra and sharp density of states. In conventional
`nonrelativistic' semiconductor systems, the natural way to realize nanoscale
quantum dots is through a confinement potential or as nanoscale islands of
semiconductor material. In both cases the quantum dots have discrete energy
spectra and electrons are localized within the quantum dot regions. In
graphene, the massless relativistic nature of the dispersion law results in
very unique properties of graphene quantum dots.  That is, the above two
approaches of realization of quantum dots have very different outcome in
graphene. While the quantum dots as isolated islands of graphene have been
successfully realized experimentally and have all the properties of
zero-dimensional systems with discrete energy spectra, the conventional
quantum dots realized through the confinement potential do not exist in
graphene.  This is due to Klein's tunneling, which provides an efficient
escape channel from a confinement potential of any strength. Therefore
electrons in graphene cannot be localized by a confinement potential.
Different approaches have been proposed to overcome this problem: generation
of an electron effective mass through interaction with a substrate,
introducing a confinement potential in a double layer system, in which
electrons have non-zero mass under applied gate voltage, or considering
special types of confinement potentials, e.g., smooth cylindrically symmetric
potentials, for which not the problem of localization but the problem of
efficient electron trapping is discussed. In Section 6, we review different
approaches to overcome the Klein's tunneling and realize graphene quantum dots
through a confinement potential.  Even for quantum dots realized as islands of
materials, the graphene islands show some unique properties. Although the main
manifestations of a two-dimensional quantization, such as the Coulomb blockade
and discrete energy spectra, are observed in experiments, the graphene
nanoscale islands show specific features. Such features are degenerate
zero-energy edge states with unique magnetic properties, specific energy level
statistics related to the Dirac billiard, and so on. These special properties
of the nanoscale graphene islands are also discussed. Finally, we also present
a brief review on quantum dots in bilayer graphene. 

In Section 7, we review the band structure of graphene nanoribbons which is
known to be modified by the presence of edges where the alteration of the
$sp^2$ network due to the mixture of the $sp$ and $sp^2$ hybridization occurs.
The nature of the zigzag edges impose localization of the electron density
with the maximum at the border carbon atoms leading to the formation of flat
conduction and valence bands near the Fermi level when the wave vector, $k\ge
2\pi/3$. The localized states are spin-polarized and in case of ordering of
the electron spin along the zigzag edges, graphene can be established in
ferromagnetic or antiferromagnetic phases. The antiferromagnetic spin ordering
of the localized states at the opposite zigzag edges breaks the sublattice
symmetry of graphene that changes its band structure and opens a gap. Because
the energetics, localization and spin ordering of the edge states can be
modified by the size of graphene nanoribbons, edge geometry, orbital
hybridization at the edges and an external electric field, their influence on
the electronic and magnetic properties of graphene are discussed.  Finally, we
turn to finite-sized bilayer graphene systems, e.g., nanoribbons, and describe
how the confinement and edge structures affect the properties of this system. 

Graphene in the real world would interact with a substrate and the surrounding
environment. If these interactions cause an imbalance of the charge or spin
distribution between graphene sublattices or modify the graphene lattice, the
sublattice or lattice symmetry of graphene will be broken, resulting in a
change of the electronic and magnetic properties of the graphene. The edges of
graphene are chemically active and prone to structural modifications and
interactions with the gas dissolved in environment, thereby influencing the
properties of graphene as well. Therefore, in Section 8, we discuss the
influence of the changes brought by the external sources into the electronic
and magnetic properties of graphene and prospects of their manipulation in a
controllable way.

\end{document}